\documentclass[onecolumn,superscriptaddress,nofootinbib,notitlepage,longbibliography,aps]{revtex4-1}
\pdfoutput=1

\usepackage{graphicx}
\usepackage{amsmath}
\usepackage{amssymb}
\usepackage{amsfonts}
\usepackage{physics}
\usepackage[dvipsnames]{xcolor}
\usepackage[linktoc=none]{hyperref}
\usepackage[utf8]{inputenc}
\usepackage{braket}
\usepackage{bbold}

\hypersetup{colorlinks,linkcolor={blue},citecolor={teal},urlcolor={violet}}

\newcommand{\documenttitle}{Electron scattering of light new particles from evaporating primordial black holes}

\newcommand{\INFN}{INFN - Sezione di Napoli, Complesso Univ. Monte S. Angelo, I-80126 Napoli, Italy}
\newcommand{\UNINA}{Dipartimento di Fisica ``Ettore Pancini'', Università degli studi di Napoli ``Federico II'', Complesso Univ. Monte S. Angelo, I-80126 Napoli, Italy}
\newcommand{\SSM}{Scuola Superiore Meridionale, Università degli studi di Napoli ``Federico II'', Largo San Marcellino 10, 80138 Napoli, Italy}
\newcommand{\NBIA}{Niels Bohr International Academy, Niels Bohr Institute, University of Copenhagen, Copenhagen, Denmark}

\begin{document}

\title{\documenttitle}

\author{Roberta Calabrese}
\affiliation{\UNINA}
\affiliation{\INFN}
\author{Marco Chianese}
\affiliation{\UNINA}
\affiliation{\INFN}
\author{Damiano F.G. Fiorillo}
\affiliation{\UNINA}
\affiliation{\INFN}
\affiliation{\NBIA}
\author{Ninetta Saviano}
\affiliation{\SSM}
\affiliation{\INFN}

\date{\today}
\begin{abstract}
Primordial black holes are a possible component of dark matter, and a most promising way of investigating them is through the product of their Hawking evaporation. As a result of this process, any species lighter than the Hawking temperature is emitted, including possible new particles beyond the Standard Model. These can then be detected in laboratory-based experiments via their interaction with the Standard Model particles. In a previous work, we have first proposed and studied this scenario in the presence of an interaction between the light new species and nucleons. Here we extend this discussion to include the case of interaction with electrons. We show that the simultaneous presence of primordial black holes and species lighter than about $100$~MeV can be constrained by the measurements of direct detection experiments, such as XENON1T, and water Cherenkov neutrino detectors, such as Super-Kamiokande. Our results provide a complementary and alternative way of investigation with respect to cosmological and collider searches.
\end{abstract}

\maketitle

\section{Introduction}

Primordial black holes (PBHs) are hypothetical black holes formed due to gravitational collapse of over-density fluctuations in the very early Universe ~\cite{Zeldovich:1967lct, Harrison:1969fb}. Different from astrophysical black holes, which must have sufficiently large masses in order to encounter the instability leading to their formation, PBHs can be produced with any mass larger than about 0.1 g \,\cite{Carr:2020gox}, because of the different equation of state involved. From the cosmological point of view, PBHs behave as a component of dark matter (DM). The possibility of observing PBHs with masses between $5 \times 10^{14}$ and $10^{18}$~g is mainly connected with the phenomenon of Hawking evaporation~\cite{hawking1974black,hawking1975particle, zeldovich1976charge, Carr:1976zz, Page:1976df, Page:1976ki, Page:1977um, PhysRevD.41.3052, MacGibbon:1991tj}, namely the emission of particles induced by the PBHs gravitational field. The  nonobservation of any production that could be connected with Hawking evaporation has allowed to severely constrain the contribution of PBHs with masses $M_{\rm PBH} \lesssim 10^{17}$~g to the dark matter content of the Universe~\cite{Calabrese:2021zfq,Coogan:2020tuf, Dasgupta:2020mqg, Wang_2021, Iguaz:2021irx}. In different mass ranges, the amount of PBHs can further be constrained by different means, see Ref.~\cite{Carr:2020gox} for a review.

Hawking evaporation produces any species that is gravitationally interacting. This includes also any elementary particle beyond the Standard Model with a mass below the Hawking temperature. In particular, if DM is dominated by a sufficiently light particle $\chi$ (with $m_\chi\lesssim 100$~MeV), it can be emitted by PBHs. In Ref.~\cite{Calabrese:2021src}, we proposed this scenario as a possible source of experimental signatures in direct detection experiments~\cite{DarkSide:2014llq, DarkSide-20k:2017zyg, Aprile_2012, XENON:2019gfn, PandaX:2021osp, LUX-ZEPLIN:2018poe,Bueno:2006re, WArP:2008mtr,ArDM:2010vgf, DEAP:2009hyz,Angloher_2005, DarkSide:2014llq, Angloher:2016ooq, DARWIN:2016hyl,SuperCDMS:2016wui,XENON:2018voc,LUX-ZEPLIN:2018poe, XENON:2019zpr,CRESST:2017ues,CRESST:2019jnq, XENON:2020kmp,COSINE-100:2021xqn}, due to the DM-nucleon interaction. We showed that the DM emitted by PBHs would provide a signal qualitatively similar to the DM boosted by cosmic-ray interactions~\cite{Cappiello:2018hsu,Bringmann:2018cvk,Ema:2018bih,Cappiello:2019qsw,Ema:2020ulo,PROSPECT:2021awi}. The  nonobservation of such a signal was used to constrain the combined parameter space of PBHs and light DM. 

Given our ignorance of the interactions of DM with the Standard Model, a natural question is what are the corresponding constraints if DM interacts only with electrons~\cite{Essig:2011nj, Essig:2012yx, Essig:2017kqs, Crisler:2018gci, DarkSide:2018ppu, DAMIC:2019dcn, Andersson:2020uwc}. Here we aim to answer this question. Recently, the same subject has been discussed in Ref.~\cite{Li:2022jxo}, where the authors apply our treatment discussed in Ref.~\cite{Calabrese:2021src} to the DM-electron interaction and obtain constraints from Super-Kamiokande~\cite{Super-Kamiokande:2017dch} and XENON1T~\cite{XENON:2020rca} measurements. However, the authors do not correctly account for the ionization of atoms that results from the DM-electron scattering. Rather, they assume electrons to be free. While this may be a reasonable approximation for Super-Kamiokande~\cite{Super-Kamiokande:2017dch,Ema:2018bih,Cappiello:2019qsw}, it is not the correct treatment for XENON1T~\cite{Lee:2015qva}. For the same reason, the discussion of energy loss in the Earth is also not applicable to the DM-electron interaction, since the DM particle does not just lose energy, but rather performs a random walk traversing the Earth for sufficiently large cross sections~\cite{Cappiello:2019qsw}. Finally, the authors of Ref.~\cite{Li:2022jxo} model the interaction between $\chi$ and electrons with a differential cross section flat in the electron recoil energy, which may not be easily realized in a realistic particle model setting.

In the present work, which was developed independently at the same time as Ref.~\cite{Li:2022jxo}, we approach the determination of the constraints with a correct treatment of the atom ionization for Xenon nuclei. We adopt an effective field theory for the DM-electron interaction, and we use it to deduce the constraints from Super-Kamiokande and XENON1T. Remarkably, the parameter space of the model is also affected by additional constraints, such as cosmological~\cite{Henning:2012rm,Sabti:2019mhn,Sabti:2021reh} and collider~\cite{Liang:2021kgw} ones, which are completely absent in Ref.~\cite{Li:2022jxo}. The constraints that we draw are not necessarily connected with the assumption that the species emitted by PBHs is dark matter. Indeed, any light species is emitted by Hawking evaporation and can subsequently be detected if it has a coupling with electrons. We emphasize that, in order to reach Earth from very far distances, such a species should be stable on cosmological scales. Therefore, throughout the paper, we will refer to this species as a generic light new particle $\chi$, and will derive the constraints with no further assumptions.

We structure the discussion as follows. In Sec.~\ref{sec:sec2} we compute the flux of the new fermionic particles $\chi$ from the evaporation of the primordial black holes. In Sec.~\ref{sec:dect} we discuss the possible detection of $\chi$ particles in two different categories of experiments: XENON1T and Super-Kamiokande. Then, we report in Sec.~\ref{sec:results} the constraints we obtain for the combined parameter space of primordial black holes and light new species. Finally, we draw our conclusions in Sec~\ref{sec:concl}.

\section{Flux of light particles from Primordial Black Holes \label{sec:sec2}}

In this section we derive the flux of a fermionic species $\chi$ emitted from evaporating PBHs. PBHs are fully characterized by mass, charge, and spin. However, evaporating, they lose mass slower than they lose charge \cite{Page:1976ki, zaumen1974upper, Carter:1974yx, Gibbons:1975kk, Page:1976df, Page:1977um}. For this reason, it is reasonable today to consider neutral PBHs that formed in the very early Universe. Moreover, we consider spinless PBHs since rotating PBHs evaporate faster causing a higher flux on Earth and leading to more stringent constraints. The radiation emitted by a single PBH of a mass ${\rm M}_{\rm PBH}$ is characterized by a thermal blackbodylike spectrum defined by the Hawking temperature $\mathcal{T}_{\rm PBH}$. In the case of nonrotating and neutral PBHs, it takes the following expression\,\cite{hawking1974black, hawking1975particle}
\begin{equation}
  k_B \mathcal{T}_{\rm PBH} = \frac{\hbar c^3}{8\pi G_{\rm N} {\rm M}_{\rm PBH}}\approx 1.06 \left[\frac{10^{16}~{\rm g}}{{\rm M}_{\rm PBH}}\right]\mathrm{MeV}\, ,
  \label{eq:Hawking_temp}
\end{equation}
where $k_B$, $G_N$ and $\hbar$ are the Boltzmann,   the gravitational and the Planck constants, respectively and $c$ is the speed of light. The differential spectrum of a single species $\chi$ from an evaporating PBH is given by~\cite{hawking1974black, hawking1975particle}
\begin{equation}
  \frac{{\rm d}N}{{\rm d}t {\rm d}T}= \frac{n_\mathrm{dof}^\chi\Gamma^\chi(T, \mathcal{T}_{\rm PBH})}{2\pi (e^{(T+m_\chi)/\mathcal{T}_{\rm PBH}} + 1)}, \,
  \label{eq:spectrum}
\end{equation}
where $T$ is kinetic energy of $\chi$, $m_\chi$ is the mass and $n_\mathrm{dof}^\chi=4$ is the number of degrees of freedom. The grey-body factor $\Gamma^\chi$ takes into account the distortions in the black-body spectrum and it is provided by the \texttt{BlackHawk} code~\cite{Arbey:2019mbc, Arbey:2021mbl}. This spectrum is peaked at an energy approximately corresponding to $\sim 5$ times the Hawking temperature and is kinematically suppressed if its mass is greater than such a value. 

For clarity of exposition, we assume the PBH mass distribution to be monochromatic. However, the method can be easily generalized to  mass distributions. We are able then to compute the $\chi$ flux from the spectrum of primordial black holes. Two components contribute to the flux, one resulting from the total emission of Extra-Galactic PBHs ($\Phi_\mathrm{EG}$) and the other from Galactic ($\Phi_\mathrm{G}$) ones. 
The Extra-Galactic contribution can be written as 
\begin{equation}
  \frac{{\rm d}\Phi_\chi^\mathrm{EG}}{{\rm d}T}=\int_{t_\mathrm{min}}^{t_{\rm max}}\dd{t} \left.\frac{\dd N}{\dd t \dd T}\right|_{E_s} \frac{f_{\rm PBH}\,\Omega_{\rm DM}\rho_\mathrm{cr}}{{\rm M}_{\rm PBH}}[1+z(t)],
  \label{eq:eg}
\end{equation}
where $z(t)$ is the redshift and $E_s= \sqrt{(E^2_\chi-m_{\rm DM}^2)(1+z(t))^2+m_{\rm DM}^2}$ is the redshifted energy of $\chi$ particles. The quantity $\Omega_{\rm DM}$ is the cosmological dark matter density and $\rho_\mathrm{cr}$ is the critical density. Indeed, since PBHs are dark matter candidates, it is convenient to refer to the PBHs abundance as the fraction of the dark matter content of the Universe:
\begin{equation}
  f_{\rm PBH} = \frac{\rho_{\rm PBH}}{\rho_\mathrm{DM}} = \frac{\Omega_{\rm PBH}}{\Omega_\mathrm{DM}}\,,
  \label{eq:f_PBH}
\end{equation}
where $f_{\rm PBH}\leq 1$. The integral in Eq.~\eqref{eq:eg} is performed from the time of matter-radiation equality ($t_\mathrm{min}$) and the age of the Universe ($t_{\rm max}$) assuming that PBHs are not fully evaporated today. We emphasize that considering times smaller than $t_{\rm min}$ would not affect the flux in a appreciable way. The Extra-Galactic component is assumed to be isotropic.

Concerning the Galactic component, this can be written as 
\begin{equation}
  \frac{\dd\Phi_\chi^\mathrm{G}}{\dd T}=\int \frac{\dd{\Omega}}{4\pi}\int_0^{l_{\rm max}}\dd{l} \frac{\dd N}{\dd t \dd T}\frac{f_{\rm PBH}\,\rho_{\rm NFW}(r(l,\phi))}{{\rm M}_{\rm PBH}},
  \label{eq:g}
\end{equation}
where $\rho_\mathrm{NFW}(r)$ is the standard Navarro-Frenk-White profile defined as~\cite{Navarro:1996gj}
\begin{equation}
  \rho_\mathrm{NFW}(r)= \rho_\odot\,\left(\frac{r_\odot}{r}\right)\left(\frac{1+r_\odot/r_s}{1+r/r_s}\right)^2
\end{equation}
where $\rho_\odot = 0.4~{\rm GeV \, cm^{-3}}$ is the local DM density, $r_\odot = 8.5~{\rm kpc}$ is the distance between the Sun and the Milky Way center, and $r_s = 20~{\rm kpc}$ is the scale radius. The galactocentric distance $r$ is
\begin{equation}
r(l,\phi)=\sqrt{r_{\odot}^2-2lr_\odot\cos \phi+l^2}\,,\hspace{1cm} l_{\rm max}=\sqrt{r_h^2-r_\odot^2 \sin^2 \phi}+r_\odot \cos \phi\,.
\end{equation}
where $r_h = 200~{\rm kpc}$ is the halo radius. We mention that since our analysis takes into account the $\chi$ flux from the whole sky (with a sizeable contribution from Extra-Galactic PBHs), the exact choice of the galactic profile is not so relevant.

According to the previous Eq.s~\eqref{eq:eg} and ~\eqref{eq:g}, the total flux of $\chi$ particles is proportional to $f_{\rm PBH}/{\rm M}_{\rm PBH}$. Therefore, the lower the PBH mass, the higher the $\chi$ flux. However, at lower masses the fraction $f_{\rm PBH}$ is typically constrained to a very small value ($<< 1$) as given in Ref.~\cite{Carr:2020gox}, thus suppressing the $\chi$ flux.

\section{Detection of light particles \label{sec:dect}}

The possibility of detecting the light particle $\chi$ is crucially dependent on its interactions with the Standard Model. In Ref.~\cite{Calabrese:2021src} we assumed that $\chi$ interacts with nucleons, whereas here we focus on its interaction with electrons. For the sake of definiteness, we consider the coupling Lagrangian
\begin{equation}
  \mathcal{L}_{\chi}= \frac{1}{\Lambda^2}\bar{\chi}\chi \,\bar{\ell} \ell\,.
  \label{eq:eff_Lagrangian}
\end{equation}
This is a non-renormalizable effective field theory (EFT) coupling characterized by the energy scale $\Lambda$, and $\ell$ refers to the Standard Model lepton fields. Such an interaction arises at the effective level from a scalar mediator with mass higher than the energies involved in the process, which in our case is at most $\sim 10$~MeV. With this condition, the scalar mediator can be integrated out, leaving the effective four-fermions interaction of Eq.~\eqref{eq:eff_Lagrangian}.

With this coupling, the $\chi$ particles emitted by PBHs can be looked for in mainly two classes of experiments:
\begin{itemize}
    \item double phase dark matter direct detection experiments based on noble liquid technology~\cite{DarkSide:2014llq, DarkSide-20k:2017zyg, Aprile_2012, XENON:2019gfn, PandaX:2021osp, LUX-ZEPLIN:2018poe,Bueno:2006re, WArP:2008mtr,ArDM:2010vgf, DEAP:2009hyz};
    \item neutrino experiments based on water Cherenkov~\cite{SNO:1987unb, ANTARES:1999fhm, Bazarko:2000id,AMANDA:2002pgr,fukuda2003super}.
\end{itemize}
In this paper, we consider XENON1T and Super-Kamiokande as representatives of the two classes, respectively. For both of them, we use the measured data to constrain the scenario of $\chi$ particles emitted from evaporating PBHs. In this section, we first present the general framework for $\chi$ interacting with the electrons bound to the atoms in the detector, and then specialize our discussion to each of the two experiments separately.

The particles $\chi$ hit detector's atoms interacting with electrons through the effective interaction of Eq.~\eqref{eq:eff_Lagrangian}. The $\chi$-$e$ scatterings would give rise to atom ionization with an outgoing non-relativistic free electron as
\begin{equation}\label{process1}
\chi+A\to \chi + A^* + e^-,    
\end{equation}
where $A$ denotes the atom, and $A^*$ is the ionized atom. The key quantity characterizing the detection of the signal is the differential event rate ${\rm d}R_\chi / {\rm d} E_r$ per unit recoil energy $E_r$, which corresponds to the free electron kinetic energy. It can be computed as:
\begin{equation}
  \frac{\dd R_\chi}{\dd E_r}= n_t\,\eta(E_r)\,F(E_r)\,\int \dd{T}\, \frac{\dd \Phi_\chi}{\dd T}\sum_{n,\,l} \frac{\dd\sigma^{n,\,l}_{\rm }}{\dd E_r}(E_r,m_\chi,T).
  \label{eq:EVENT_RATE}
\end{equation}
Here, the quantity $n_t$ is the number of detector's targets per tonne and $\eta$ is the detector's efficiency. The quantity $F(E_r)$ is the Fermi factor that takes into account the distortion of the scattered electron wavefunction by the presence of the atom. In the non-relativistic limit, we have
\begin{equation}
  F(E_r) = \frac{2\pi \nu}{1-e^{-2\pi \nu}},
  \label{eq:Fermi_factor}
\end{equation}
where $\nu = Z_{\rm eff}(\alpha m_e/\sqrt{2\,m_e\,E_r})$ and $Z_{\rm eff}$ is the effective charge that is felt by the scattered electron. In our analysis, we conservatively set $Z_{\rm eff}=1$.\footnote{In general $Z_{\rm eff}$ is greater than one since the shielding of the escaping electron by the remaining bounded electrons is imperfect. In Ref.\,\cite{Lee:2013wza}, it was pointed out that $Z_{\rm eff}= 1$ is a good approximation for outer-shell electrons. Moreover, assuming $Z_{\rm eff}$ greater than the unity would enhance the event rate. Therefore, our choice is good for outer-shell electrons and conservative for the others.} Finally, $\dd\Phi_\chi/\dd T$ is the differential $\chi$ flux introduced in Sec.~\ref{sec:sec2}, and $\dd\sigma^{n,l}/\dd E_r$ is the differential cross section for scattering of a $\chi$ particle on a bound electron with principal quantum number $n$, orbital quantum number $l$, and recoil energy $E_r$. Denoting by $E_b^{n,\,l}$ the electron binding energy of the atomic orbital $(n,l)$, from energy conservation we have
\begin{equation}
    E_\chi - |E_b^{n,\,l}|=  E_\chi^\prime +  E_r
    \label{Eq:Energy_conservation}
\end{equation}
where $E_\chi = T+m_\chi$ and $E_\chi^\prime$ are the initial and final energy of $\chi$ particle, respectively.

The strength of the $\chi$-$e$ interaction can be parameterized in terms of the coupling at the Lagrangian level. However, it is more common to express it in terms of the cross section on a free electron at a fixed momentum transfer $\alpha m_e$
\begin{equation}
    \overline{\sigma}_{\chi e} = \frac{\mu_{\chi e}^2}{\pi\, \Lambda^4} \left(1 + \frac{\alpha^2\,m_e^2}{4\,m_\chi^2}\right)\,
    \label{eq:cross_ref}
\end{equation}
with $\alpha$ being the fine-structure constant. This is the quantity that we aim to constrain. Before discussing how the analysis proceeds for the two experiments separately, it is worth noticing that the same $\chi$-$e$ interaction would cause an attenuation effect of the $\chi$ flux due to the propagation in the atmosphere and the Earth~\cite{Starkman:1990nj, Mack:2007xj, Kavanagh:2016pyr, Emken:2018run}. However, as shown by Ref.~\cite{Ema:2018bih,Cappiello:2019qsw} the attenuation is negligible for $\overline{\sigma}_{\chi e} \lesssim 10^{-31}~{\rm cm^2}$. For this reason, we restrict our analysis to smaller cross sections only.

\subsection{XENON1T event rate}
\begin{figure}[t!]
  \centering
  \includegraphics[width=0.75\textwidth]{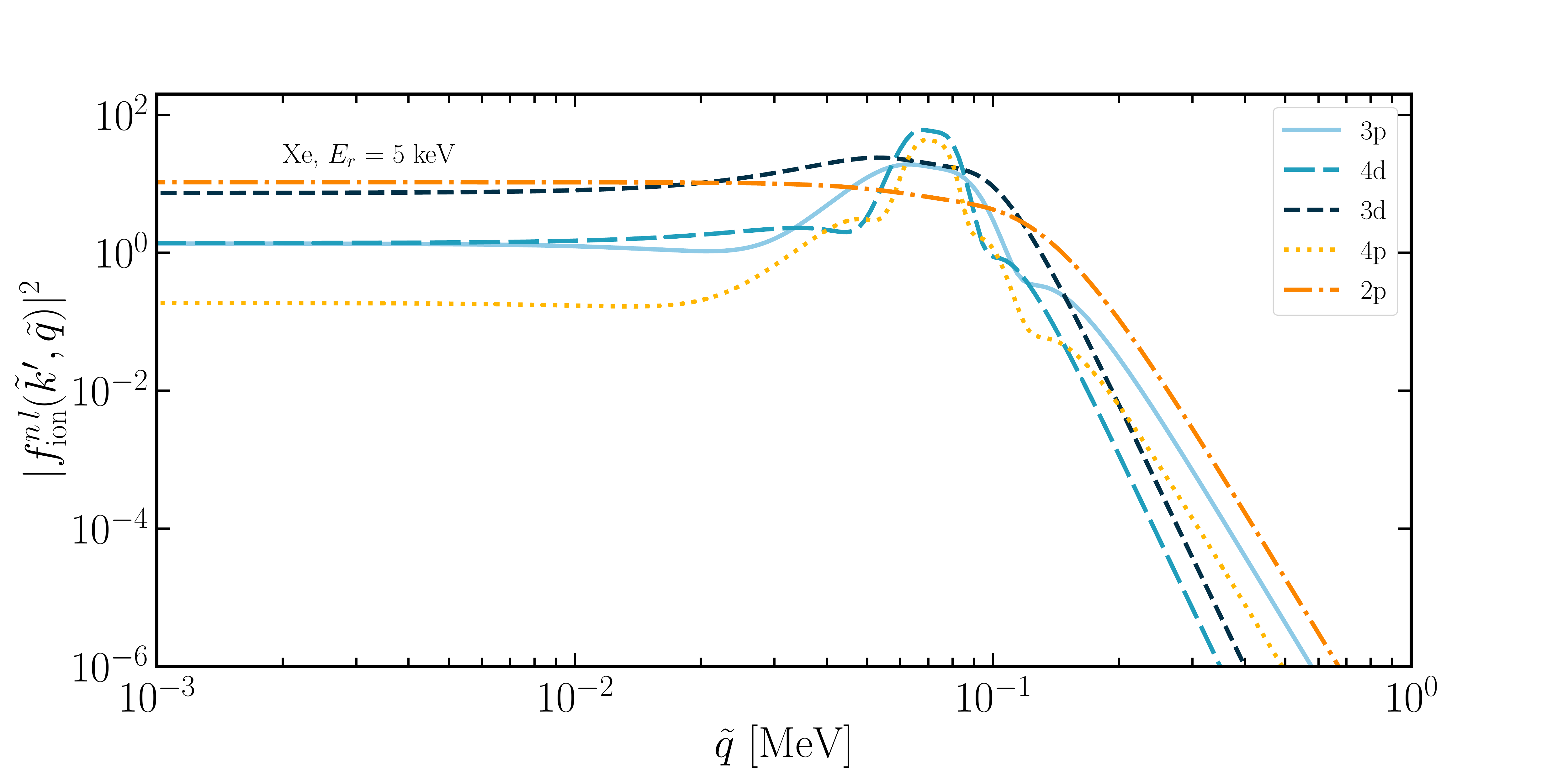}
  \caption{Xenon ionization function as a function of the momentum transfer $\tilde{q}$ for  orbitals and with $E_r = 5$ keV and $\tilde{k}' = \sqrt{2\,m_e\,E_r}$.}
  \label{fig:ionization_function}
\end{figure}

XENON1T utilizes a dual-phase liquid Xe time projection chamber, used in most of the experiments based on noble liquid, with a 2.0-tonne active target. Each interaction inside the detector produces a prompt scintillation signal (signal 1) and a delayed electroluminescence signal (signal 2). The technology of this kind of detector allows one to study both the elastic scattering between $\chi$ particles and nuclei, and the ionization of electrons bound to atoms due to $\chi$-$e$ interactions. Here, we use the measurements from XENON1T~\cite{XENON:2020rca} of the event rate for recoil energies $E_r$ in the range $[1-30]$~keV to constrain the $\chi$-$e$ cross section in Eq.~\eqref{eq:cross_ref}. For this experiment the target density is $n_t = 4.59\cdot10^{27}\,{\rm ton^{-1}}$ corresponding to the number of detector's Xe nuclei, and we use the detector efficiency $\eta(E_r)$ reported in Ref.~\cite{XENON:2020rca}.

Assuming the outgoing electron to be non-relativistic, it is possible to show that the differential cross section takes the following expression\footnote{In this paper, given a four-momentum $k$ we denote $\tilde{k} = |\boldsymbol{k}|$.}
\begin{equation}
     \frac{d\sigma^{n,\,l}}{dE_r}(E_r,m_\chi,T)=\frac{1}{8\pi\,\Lambda^4\,\tilde{p}^2\,m_e} \int_{\tilde{q}_-}^{\tilde{q}_+} \dd{\tilde q} \tilde{q} \frac{1}{2\,\tilde{k}^{\prime\,2}}\left|f^{n,\,l}_{\rm ion}(\tilde{k}^\prime, \tilde{q})\right|^2 \left[(p\cdot(p-q)+m_\chi^2)\,(k^\prime\cdot(k^\prime - q)+m_e^2)\right]\,.
    \label{eqsigmanl1}
\end{equation}
where $p$ is the four-momentum of incoming $\chi$ particles, $k'$ is the one of the outgoing free electron, and $q$ is the momentum transfer. The integration extremes $\tilde{q}_\pm$ are the minimum and maximum momentum transfer kinematically allowed:
\begin{equation}
  \tilde{q}_\pm = \sqrt{(T+m_\chi)^2-m_\chi^2}\pm \sqrt{\left(T+m_\chi -\varepsilon_{n,l} \right)^2-m_\chi^2}\,,
  \label{eq:q_min_&_max}
\end{equation}
where $\varepsilon_{n,l}=E_r+|E_b^{n,l}|$.
Moreover, the quantities $f^{n,l}_{\mathrm{ion}}$ are the ionization functions, which can be written as~\cite{Kopp:2009et,Lee:2015qva,Catena:2019gfa}
\begin{equation}
  |f^{n,\,l}_\mathrm{ion}(\tilde{k}^\prime, \tilde{q})|^2=\frac{(2l+1)\tilde{k}^{\prime\,2}}{4\pi^3\tilde{q}}\int_{\tilde{k}^\prime-\tilde{q}}^{\tilde{k}^\prime + \tilde{q}} \dd{\tilde{k}}\,\tilde{k} |\widetilde{\mathcal{R}}_{n,\,l}(\tilde{k})|^2\,,
  \label{eq:ionization_function_2}
\end{equation}
where the integral is performed for all the allowed values for the module of the electron initial momentum, $\tilde{k}$, and $\widetilde{\mathcal{R}}_{n,\,l}(\tilde{k})$ is the Fourier transform of the radial part of the bound electron wavefunction (see~\cite{Kopp:2009et} for details)
{\small
\begin{equation}
\widetilde{\mathcal{R}}_{n,\,l}(\tilde{k})= \sum_\theta \frac{C_{nl\theta}2^{-l+n_{l\theta}}}{\Gamma(3/2+l)}
  \left(\frac{2\pi a_0}{Z_{l\theta}}\right)^{3/2}
  \left(\frac{i\tilde{k}a_0}{Z_{l\theta}}\right)^l
  \frac{(1+n_{l\theta}+l)!}{\sqrt{(2n_{l\theta})!}}\, 
  _2F_1\left[\frac{(2+n_{l\theta}+l)}{2}, \frac{(3+n_{l\theta}+l)}{2}, \frac{3}{2}+l, -\left(\frac{\tilde{k}a_0}{Z_{l\theta}}\right)^2\right]
  \label{eq:chi(k)_1}
\end{equation}}
where $_2F_1(a,b,c,x)$ is the hyper-geometric function. All the coefficients appearing in this expression, $C_{nl\theta}$, $n_{l\theta}$ and $Z_{l\theta}$, are tabulated in Ref.\,\cite{BUNGE1993113}, while $a_0$ is the Bohr radius. In Fig.~\ref{fig:ionization_function} we show some examples of the ionization functions for  Xenon orbitals as a function of the momentum transfer $\tilde{q}$ once the electron recoil energy is fixed to 5 keV. 

Expanding the terms in the last square bracket in Eq.~\eqref{eqsigmanl1} in the non-relativistic regime for the electron and substituting the reference cross section defined in Eq.~\eqref{eq:cross_ref}, we obtain
\begin{equation}
    \frac{\dd\sigma^{n,\,l}}{\dd E_r}(E_r,m_\chi,T)= \frac{ \overline{\sigma}_{\chi e}\,m_e}{8\,\mu_{\chi e}^2 \tilde{k}^{\prime \,2} \, \tilde{p}^2 \, \left(1+\frac{\alpha^2 \,m_e^2}{4\,m_\chi^2}\right)}\int_{\tilde{q}_-}^{\tilde{q}_+}\dd{\tilde{q}} \tilde{q} \left|f_{ion}^{n,\,l}(\tilde{q}, \tilde{k}^\prime)\right|^2 \left(2m_\chi^2 + \frac{\tilde{q}^2-\varepsilon^2_{n,l}}{2} \right),
     \label{eq:XENON}
\end{equation}

Having determined the cross section, we proceed to obtain the event rate in the XENON1T detector and compare it with the measurements from the collaboration. The free parameters of the model are ${\rm M}_{\rm PBH}$, $f_{\rm PBH}$, $m_\chi$, and $\overline{\sigma}_{\chi e}$. Since our aim is to constrain the model, we adopt the statistical procedure suggested in Ref.~\cite{Cowan:2010js} for setting upper limits. We define the chi-squared variable
\begin{equation}
    \chi^2\left({\rm M}_{\rm PBH},\, m_\chi,\, \overline{\sigma}_{\chi e},\, f_{\rm PBH}\right) := \sum_i \frac{\left[\frac{dR_{\rm obs}}{dE_r}-\left(\frac{dR_{\rm BCK}}{dE_r}+\frac{dR_\chi}{dE_r}\right)\right]^2_{E_r = E_r^i}}{\sigma_i^2}\,.
    \label{eq:chi_square}
\end{equation}
Here $dR_{\rm obs}/dE_r$ is the observed event rate, while $dR_{\rm BCK}/dE_r$ is the estimated background event rate. For both of them we take the results of Ref.\,\cite{XENON:2020rca}. The sum is performed over all the energy bins, and  $\sigma_i$ are the uncertainties on $dR_{\rm obs}/dE_r(E_r^i)$. We then set the test statistic $\lambda$ for upper limits as
\begin{equation}
    \lambda=\begin{cases}
    \chi^2\left({\rm M}_{\rm PBH},\, m_\chi,\, \overline{\sigma}_{\chi e},\, f_{\rm PBH}\right)-\chi^2\left({\rm M}_{\rm PBH},\, m_\chi,\, \hat{\overline{\sigma}}_{\chi e},\, f_{\rm PBH}\right),\qquad \overline{\sigma}_{\chi e}>\hat{\overline{\sigma}}_{\chi e}\\
    0,\qquad \overline{\sigma}_{\chi e}<\hat{\overline{\sigma}}_{\chi e}\,, 
    \end{cases}
\end{equation}
where $\hat{\overline{\sigma}}_{\chi e}$ is the value of the cross section which minimizes the chi-squared. In this way, we provide the most conservative upper limits on this quantity. We then exclude at $90\%$ confidence level the region of the parameter space in which $\lambda>2.71$, following the prescription in Ref.~\cite{Cowan:2010js}.

\subsection{Super-Kamiokande's event rate}

Super-Kamiokande is a water Cherenkov detector realized by a 
cylindrical tank filled with 50kt of water.
Charged particles in the water produce Cherenkov radiation that is recorded by the photo-multipliers.
The light particle $\chi$ is also expected to yield Cherenkov radiation when scattering inside the detector. Super-Kamiokande has observed $N_{\rm SK} = 4042$ events in the recoil energy range between $[ 0.1 - 1.33]~{\rm GeV}$~\cite{Super-Kamiokande:2017dch}.

In the case of Super-Kamiokande, previous analyses assume the electrons to be free and at rest in the observer frame~\cite{Necib:2016aez, Ema:2018bih, Cho:2020mnc, Granelli:2022ysi}. In this work, we follow the same approach and assume the number of targets to be $n_t =3.34 \times 10^{28}~{\rm ton^{-1}}$ and the detector efficiency to be equal to 0.93 in the energy range considered~\cite{Super-Kamiokande:2017dch}. In the limit of free electrons, we have $F(E_r) = 1$, and the differential cross section is simply given by
\begin{equation}
    \frac{\dd\sigma}{\dd E_r} = \frac{\overline{\sigma}_{\chi e}\,\Theta(E_{\rm max}-E_r)}{8\,\mu^2_{\chi e}\, \tilde{p}^2\left(1+\frac{\alpha^2 \,m_e^2}{4\,m_\chi^2}\right)} (2m_e+E_r)(2m_\chi^2 + m_e E_r)
\end{equation}
where $E_{\rm max}$ is the maximum allowed recoil energy equal to
\begin{equation}
    E_{\rm max} = \frac{2\,m_e\, T\,(T+2\,m_\chi)}{\left(\left(m_e+m_\chi\right)^2+2\,m_e\,T\right)}\,.
\end{equation}
Plugging this equation into Eq.~\eqref{eq:EVENT_RATE} we can compute the expected number of events in Super-Kamiokande provided by $\chi$ particles. Following Ref.~\cite{Ema:2018bih}, we conservatively obtain the constraints on $\overline{\sigma}_{\chi e}$ by simply requiring that
\begin{equation}
    \mathcal{E}_{\rm SK} \times \int_{0.1 {\rm GeV}}^{1.33 {\rm GeV}}\dd{E_r}\,\frac{\dd R_\chi}{\dd E_r} < N_{\rm SK}\,,
\end{equation}
where $\mathcal{E}_{\rm SK} = 161.9~{\rm kton~yr}$ is the Super-Kamiokande exposure~\cite{Super-Kamiokande:2017dch}.

\section{Results \label{sec:results}}

\begin{figure}[t!]
  \centering
  \includegraphics[width=0.47\textwidth]{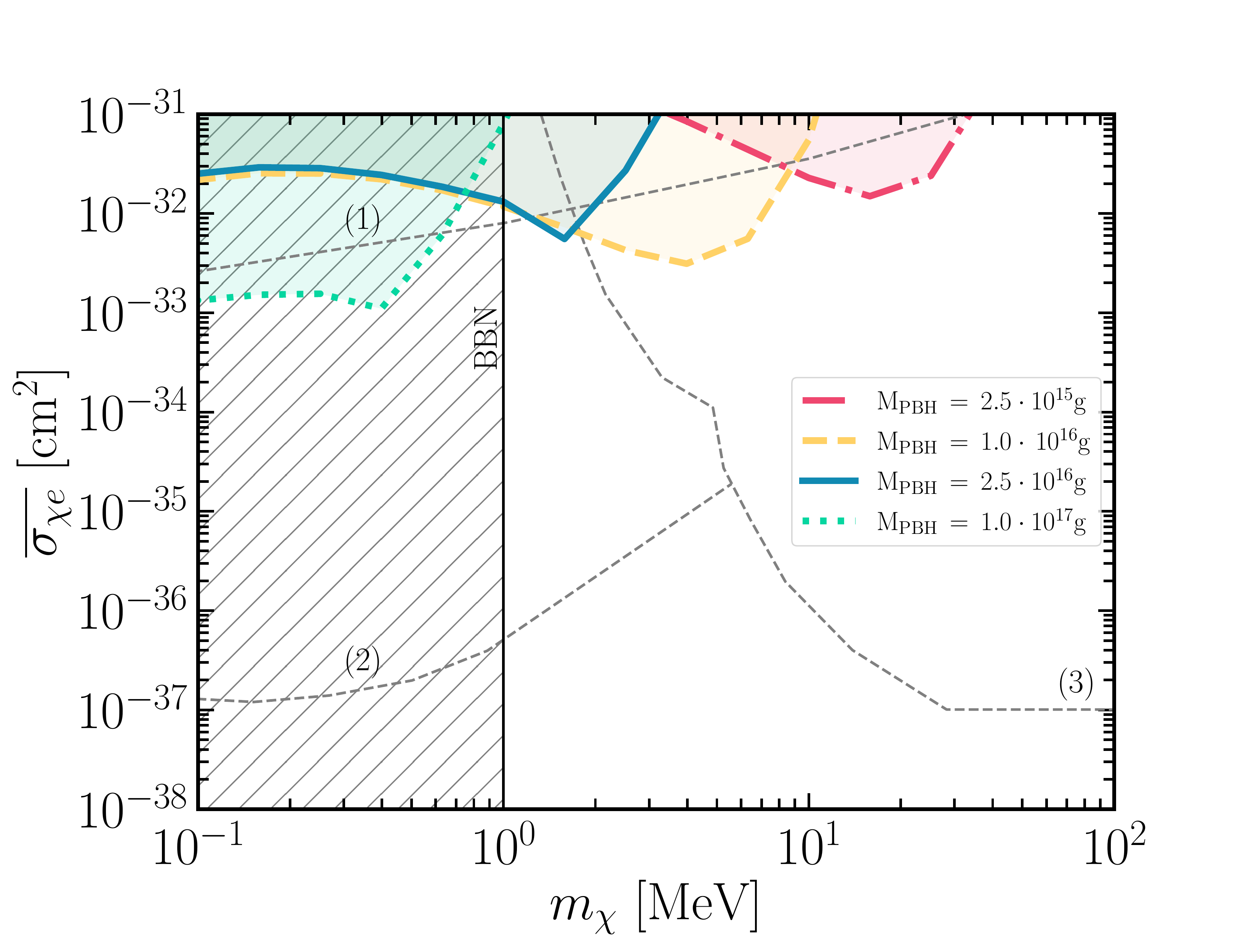}
  \includegraphics[width=0.47\textwidth]{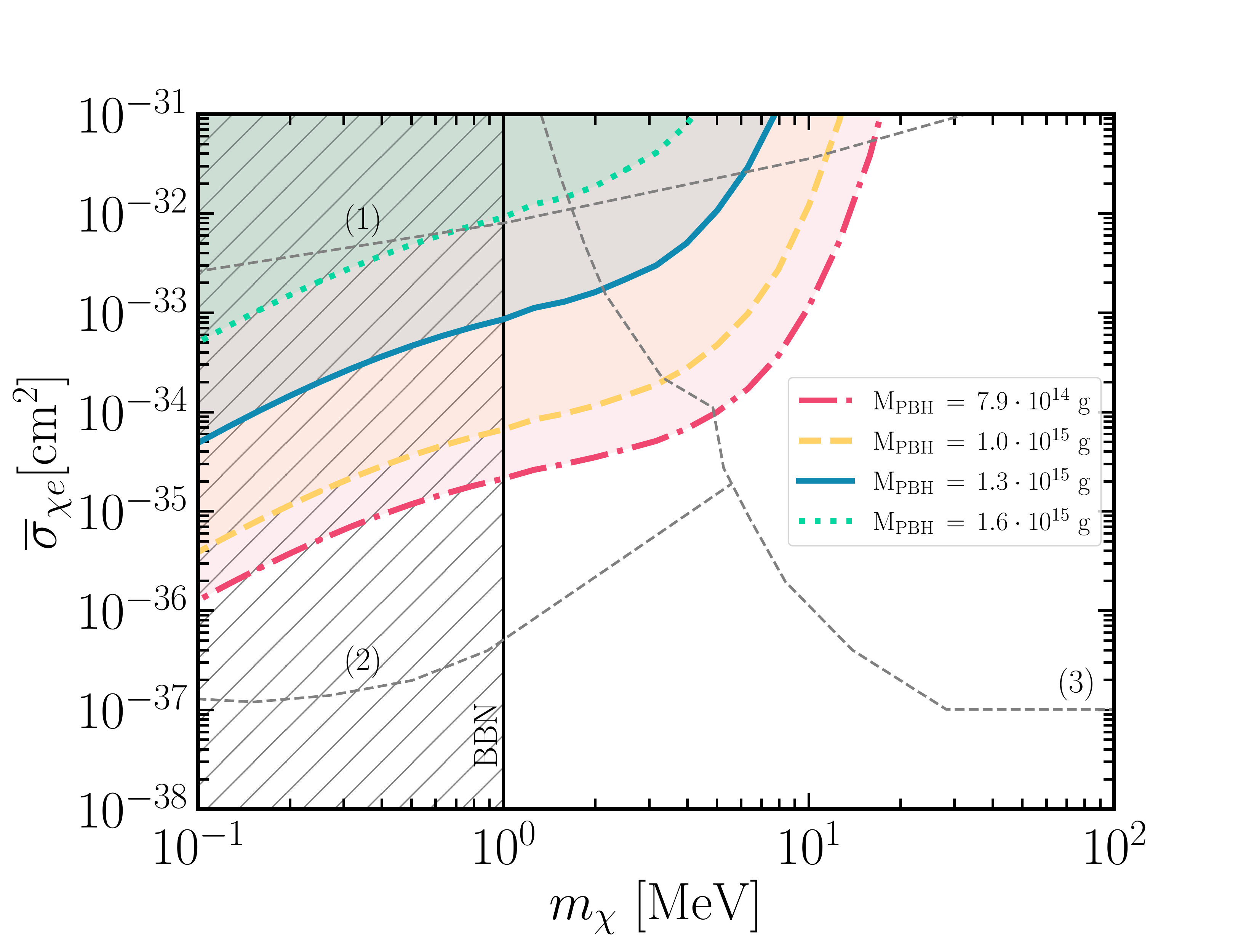}
  \caption{Constraints on the reference $\chi$-$e$ cross section $\overline{\sigma}_{\chi e}$ in Eq.~\eqref{eq:cross_ref} as a function of $\chi$ mass $m_\chi$ in case of XENON1T (left panel) and Super-Kamiokande (right panel). The color shaded regions represents our results for  PBH masses once their fraction $f_{\rm PBH}$ is fixed to the corresponding maximum value according to current constraints (see footnote 3). The hatched region is excluded by Big Bang Nucleosynthesis, while the thin dashed lines correspond to the limits that apply only if $\chi$ particles are dark matter: (1) boosted dark matter from cosmic-ray up-scatterings~\cite{Cappiello:2019qsw};
  (2) Solar reflection with XENON1T~\cite{An:2017ojc};
  (3) combined constraints from dark matter direct detection experiments (see Ref.~\cite{Cappiello:2019qsw}).}
  \label{fig:Constraints_1}
\end{figure}

The analysis at XENON1T~\cite{XENON:2020rca} and Super-Kamiokande~\cite{Super-Kamiokande:2017dch} allows to constrain the combined parameter space of the model, which is expressed in terms of the PBH mass $M_{\mathrm{PBH}}$, the PBH fraction of DM $f_\mathrm{PBH}$, the mass of the light particle $m_\chi$, and the cross section $\overline{\sigma}_{\chi e}$ introduced above in Eq.~\eqref{eq:cross_ref}. The large dimensionality of the parameter space requires to show the constraints as sections in the two-dimensional planes. In particular, in Fig.~\ref{fig:Constraints_1} we show the constraints in the $m_\chi-\overline{\sigma}_{\chi e}$ plane for varying masses of the PBHs. For each mass of the PBHs, we have assumed $f_{\mathrm{PBH}}$ to be as large as allowed by the present constraints on the PBH parameter space alone. We present separately the constraints from XENON1T (left panel) and Super-Kamiokande (right panel). We show as well the constraints (dashed thin lines) which would apply if $\chi$ is identified as the dominant component of dark matter (see the caption).

Due to the differences in the cuts in the recoil energies, the two experiments are able to probe a different range of PBH masses. In particular, XENON1T can look at somewhat heavier PBHs with a smaller Hawking temperature and corresponding smaller recoil energies. We find that in both cases the constraints that can be drawn, assuming that PBHs are as many as allowed,\footnote{In this analysis, we considering the constraints on $f_{\rm PBH}$ derived from Extra-Galactic gamma-ray data~\cite{Carr:2020gox} and from isotropic X-ray observations~\cite{Iguaz:2021irx}. The latter provides the stronger limit for $\mathrm{M}_\mathrm{PBH} \gtrsim 10^{16}~{\rm g}$. Such constraints do not depend on the choice of the dark matter galactic profile since they are based on Extra-Galactic and isotropic measurements.} are complementary to the constraints on light DM. However, if the species $\chi$ is not identified as DM, the constraints from the dashed lines do not apply. Nevertheless, for $m_\chi \lesssim 1$~MeV any light species sufficiently strongly interacting would be in serious tension with a successful Big Bang Nucleosynthesis. For this reason, we consider this region as completely excluded, even though specific models can be devised to evade these constraints~\cite{Elor:2021swj}.

Among the possible additional constraints, we can consider the ones from collider experiments~\cite{Liang:2021kgw}, which we do not show for purely graphical reasons, since they would exclude the region $\overline{\sigma}_{\chi e}\gtrsim 10^{-44}$~cm$^2$, which is much lower than the range of our figures. However, we have to stress that the collider constraints are strongly sensitive to the mediator mass, as shown in Ref.~\cite{Liang:2021kgw}. While the authors do not provide the constraints for a scalar mediator with arbitrary mass, which we are studying in this work, for the vector mediator model they clearly show that the constraints significantly weaken for lighter mediators, already at a mass of $0.7$~GeV. Furthermore, the constraints from collider are expected to be limited from above by a ceiling, when the cross section becomes sufficiently large for the $\chi$ particle to interact in the calorimeter. This has been explicitly studied for the case of $\chi$-nucleon interaction~\cite{Cappiello:2018hsu}, yet no such study exists for $\chi$-electron interaction. Therefore, obtaining constraints which are complementary to the collider ones is especially necessary. Finally, light dark matter coupled to electrons can be emitted in dense astrophysical environments, in particular in supernova for these  DM masses. Indeed, the emission of light particles from supernova would lead to an additional cooling mechanism and change the duration of the neutrino burst. This would have led to observable consequences for the case of SN1987A. The constraints from SN1987A are discussed in the context of a specific interaction model with vector mediator in Ref.~\cite{Chang:2018rso}. These results show that, at $\overline{\sigma}_{\chi e}\gtrsim 10^{-39}$~cm$^2$, SN1987A does not lead to significant constraints because the $\chi$ particle would be trapped inside the supernova without leading to an observable signal. Since we focus on a scalar interaction, these results are not directly applicable. However, they suggest that at the large cross sections we consider, the constraints from SN1987A would not be applicable.

A crucial point is the comparison between our results and the results of Ref.~\cite{Li:2022jxo}, since they are representing the same constraints. For Super-Kamiokande we find constraints which, for the same PBH mass, are about two orders of magnitude stronger. This is mainly due to our use of a concrete particle model for the $\chi$-electron interaction, whereas Ref.~\cite{Li:2022jxo} simply assumes a differential cross section flat in the electron recoil energy, which is not necessarily realizable in a realistic particle model. For XENON1T the constraints we find are significantly weaker, by as much as three orders of magnitude. We attribute this difference mostly to the complete different treatment of the DM-electron interaction. In fact, whereas in Ref.~\cite{Li:2022jxo} the electrons are treated as free, we correctly account for the ionization of the xenon atoms in the detector. 

\begin{figure}[t!]
  \centering
  \includegraphics[width=0.75\textwidth]{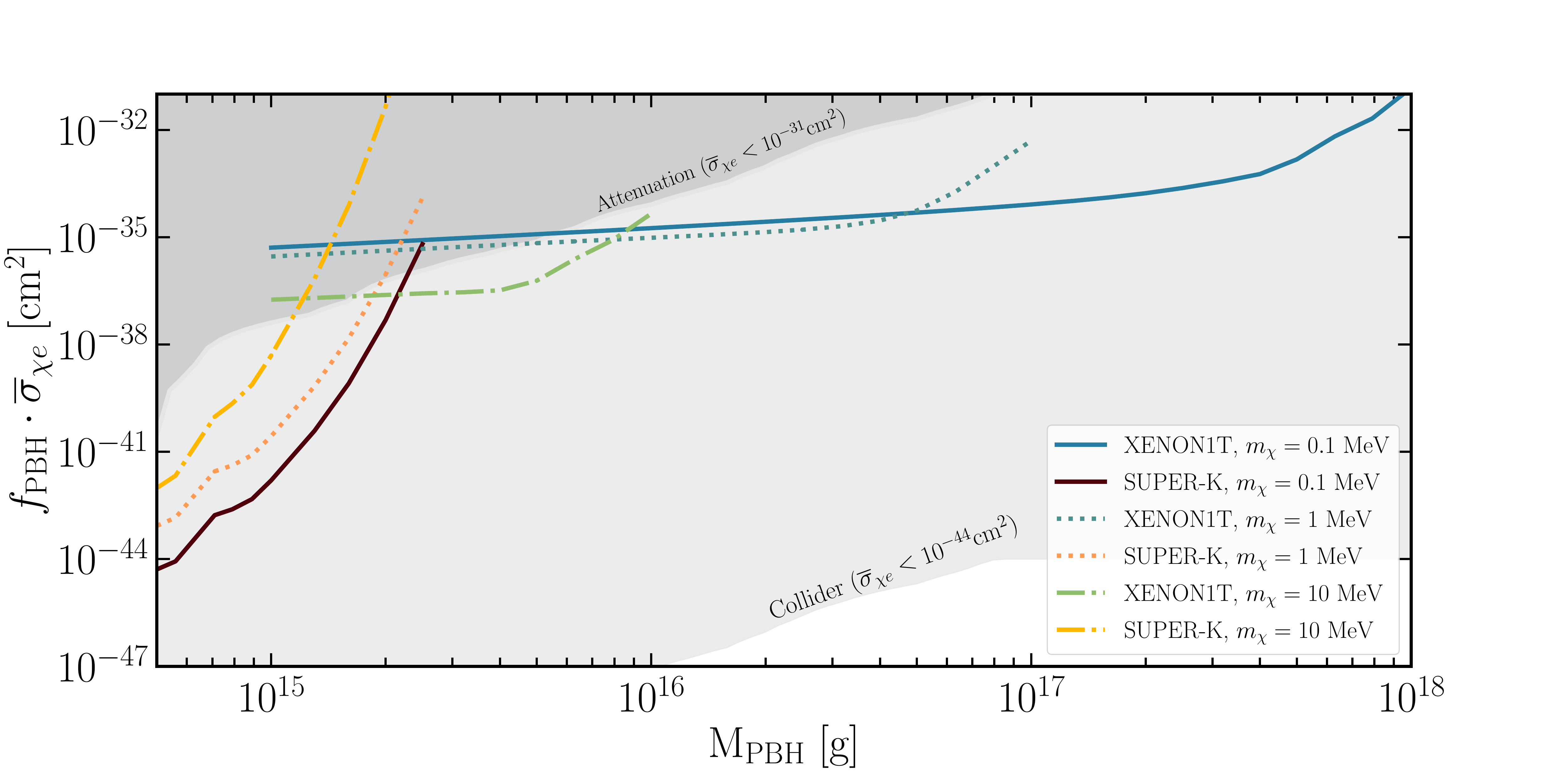}
  \caption{Constraints shown in the plane $(f_{\rm PBH}\cdot \overline{\sigma}_{\chi e})-{\rm M}_{\rm PBH}$ for different masses $m_\chi$. In warm tones are shown the constraints obtained using Super-Kamiokande, while in cold tones are shown the constraints obtained using XENON1T. The dark grey band shows the region that cannot be probed in our analysis due to existing constraints on $f_{\rm PBH}$ (see footnote 3) and assuming $\overline{\sigma}_{\chi e} \lesssim 10^{-31}~{\rm cm^2}$ to neglect attenuation effects. The light grey band is excluded by requiring $\overline{\sigma}_{\chi e} \lesssim 10^{-44}~{\rm cm^2}$ according to collider constraints~\cite{Liang:2021kgw}.}
  \label{fig:Constraints_2}
\end{figure}

A more compact, although less conventional, way of showing our results is to exploit the proportionality of the signal in any detector to the product $f_{\mathrm{PBH}} \cdot \overline{\sigma}_{\chi e}$. Therefore, the parameters effectively determining the constraints are in reality three, namely $m_\chi$, $\mathrm{M}_{\mathrm{PBH}}$, and $f_{\mathrm{PBH}} \cdot \overline{\sigma}_{\chi e}$. We show the constraints in the $(f_{\mathrm{PBH}} \cdot \overline{\sigma}_{\chi e}) - \mathrm{M}_{\mathrm{PBH}}$ plane in Fig.~\ref{fig:Constraints_2} both for XENON1T and Super-Kamiokande. This figure again shows that the range of PBH masses probed by the two experiments is different. Furthermore, for XENON1T this range potentially extends up to $10^{18}$~g depending on the mass $m_\chi$, since the light particle is only emitted if $m_\chi$ is smaller than the Hawking temperature of the PBH.

\section{Conclusions \label{sec:concl}}

In this paper, we have extended the work~\cite{Calabrese:2021src} where we proposed to test the potential existence of new light particles beyond the Standard Model through the evaporation of primordial black holes. In particular, we have focused on the possible interactions of such new particles with the electrons through an effective coupling mediated by a heavy scalar. Hence, we have investigated the consequent detection of $\chi$ particles from evaporating primordial black holes in dark matter direct detection experiments, such as XENON1T, and neutrino detectors, such as Super-Kamiokande. The  nonobservation of the expected signal has been used to set constraints on the combined parameter space of primordial black holes and light new particles. In the case of XENON1T, we have performed a binned likelihood analysis comparing the observed event rate with the one predicted in our scenario. We have properly taken into account the ionization process of Xe atoms due to $\chi$-$e$ scattering. On the other hand, for Super-Kamiokande we have performed a more conservative analysis based on the total number of events detected. We have found that XENON1T (Super-Kamiokande) is able to constrain the $\chi$-$e$ cross section down to $10^{-32}~{\rm cm^2}$ ($10^{-35}~{\rm cm^2}$) for $m_\chi = 1~{\rm MeV}$ in case of PBH with masses from $5 \times 10^{14}$ to $10^{18}$~g. These limits are complementary to cosmological and collider constraints and do not require the $\chi$ particles to be dark matter.

\subsection*{Acknowledgements}
We thank Gennaro Miele and Stefano Morisi for useful comments and discussions. This work was supported by the research grant number 2017W4HA7S ``NAT-NET: Neutrino and Astroparticle Theory Network'' under the program PRIN 2017 funded by the Italian Ministero dell'Universit\`a e della Ricerca (MUR) and by the research project TAsP (Theoretical Astroparticle Physics) funded by the Istituto Nazionale di Fisica Nucleare (INFN). RC acknowledges also financial support from the agreement ASI-INAF n.2017-14-H.O.. The work of DFGF is partially supported by the {\sc Villum Fonden} under project no.~29388.   This project has received funding from the European Union's Horizon 2020 research and innovation program under the Marie Sklodowska-Curie grant agreement No.~847523 ‘INTERACTIONS’.

\bibliography{references}
\end{document}